\documentclass{aa}
\usepackage{graphicx}

\usepackage[colorlinks=true, allcolors=blue]{hyperref}

\newcommand{\re}[1]{\textcolor{red}{(ref) } }
\newcommand{\new}[1]{#1}
\newcommand{\neww}[1]{#1}
\newcommand{\nw}[1]{#1}
\begin{document}
%\interfootnotelinepenalty=1000

\title{Probing binary neutron star mergers in dense environments using afterglow counterparts}

\author{Rapha\"{e}l Duque$^{1}$\thanks{E-mail: duque@iap.fr},
Paz Beniamini$^{2}$,
Fr\'{e}d\'{e}ric Daigne$^{1}$
and Robert Mochkovitch$^{1}$}

\institute{$^{1}$Sorbonne Universit\'{e}, CNRS, UMR 7095, Institut d'Astrophysique de Paris, 98 bis boulevard Arago, 75014 Paris, France\\
$^{2}$Division of Physics, Mathematics and Astronomy, California Institute of Technology, Pasadena, CA 91125, USA}
\authorrunning{Duque et al.}
\titlerunning{Probing mergers in dense media using afterglows}

%\institute{Sorbonne Université, CNRS, UMR 7095, Institut d’Astrophysique de Paris, 98 bis boulevard Arago, 75014 Paris, France}

\abstract{The only binary neutron star merger gravitational wave event with detected electromagnetic counterparts \nw{recorded to date is GRB170817A. This merger occurred} in a rarefied medium with a density smaller than $10^{-3}-10^{-2}~{\rm cm}^{-3}$. Since kicks are imparted to neutron star binaries upon formation, and due to their long delay times before merger, such low-density circum-merger media are generally expected. However, there is some indirect evidence for fast-merging or low-kick binaries, which would coalesce in denser environments. Nonetheless, present astronomical data are largely inconclusive on the possibility of these high-density mergers. We describe a method to directly probe this hypothetical population of high-density mergers through multi-messenger observations of binary neutron star merger afterglows, exploiting the high sensitivity of \new{these signals to the density of the merger environment}. This method is based on a sample of merger afterglows that has yet to be collected. Its constraining power is large, even with a small sample of events. We discuss the method's limitations and applicability. In the upcoming era of third-generation gravitational wave detectors, this method's potential will be fully realized as it will allow \neww{us} to probe mergers that occurred soon after the peak of cosmic star formation, \new{provided the follow-up campaigns are able to locate the sources}.}


\maketitle

\section{Introduction}
\label{sec:intro}
%The distribution of delay-times between the formation and merger of binary neutron stars (BNS) remains an open question. 
\new{Upon the second supernova leading to their formation, binary neutron stars (BNS) are kicked away from their dense star-forming birth regions \citep{Bl61, Bo61, FK97}, allowing them to migrate to a different environment before merging \citep{PZY98}. Because of the slow rate of orbital decay, this migration is generally expected to be long enough for the merger to occur in a rarefied medium \citep[e.g.][]{B99}.} \new{This was the case for the up-to-now single BNS merger gravitational wave (GW) event with afterglow counterpart, GRB170817A, which occurred in a medium with density $n \lesssim 10^{-2}~{\rm cm}^{-3}$ (\citealt{hallinan17,hajela19}, and see Sect.~\ref{sec:3.1} below).}

\new{Supernova kicks are poorly constrained in the general picture \citep{PPR} and may be variable from a system to another \citep[e.g.,][]{P+04}. Overall, they define the system's velocity during migration and affect its initial separation and eccentricity and therefore its merger time \citep{BP95, K96, BB99}. For systems with the lowest kick velocities or shortest delay-times, we expect the distances covered during migration to be shorter than for the rest of the population, leading to the possibility of binaries merging in environments that are much denser than those encountered by systems with long migrations. We refer to these events, with densities $n \gtrsim 1~{\rm cm}^{-3}$, as `high-density mergers'.}

\new{Furthermore, there exist some theoretical stellar evolutionary pathways leading to short-delay or low-kick systems and therefore high-density mergers \citep[e.g.,][]{ivanova03, 181, secunda+19}. Whether these mechanisms are realized is not certain and, as we show below, current data is inconclusive regarding the \new{importance} of this class of mergers.}

\new{In the electromagnetic domain, modelling of the afterglows of short gamma-ray bursts (GRB) is one probe of the environments of BNS mergers \citep[e.g.,][]{118}. Unfortunately, because of the poor localization of most short GRBs and of the relative faintness of their afterglows, the X-ray afterglow of only a small fraction have been found, and less than a handful have detected afterglows in the radio band \citep{davanzo15}.}

% occurred in a medium typical of early-type galaxies. Indeed, the circum-merger density was . Throughout this work, we will use `dense' to refer to media with densities significantly larger than this, i.e., with $n \gtrsim 1~{\rm cm}^{-3}$.

In recent works, \citet{145} and \citet{185} have studied the afterglows expected as counterparts to GW signals from BNS mergers in present and future observing runs. Starting from a population model motivated by knowledge of short GRB, \citet{185} have found that the fraction of afterglows detectable in the radio band sharply increases with the density $n$ of the medium hosting the mergers. This is due to the fact that (i) radio frequencies $\nu_{R}$ are expected to fall between the injection and cooling frequencies $\nu_i$ and $\nu_c$ of the synchrotron slow-cooling regime for the bulk of the population, and (ii) in this regime, the afterglow peak flux scales as $F_p \propto n^{\frac{p+1}{4}}$ (e.g., \citealt{114}), where $p \sim 2-3$ is the spectral index of the non-thermal electron population accelerated at the jet's forward shock front. Thus, should there be mergers in high-density environments, these would be over-represented in the afterglow population with respect to their actual number. In other words, the radio afterglow acts as an \new{amplifier} for these higher density mergers.

\nw{Given a statistical flux-limited sample of BNS merger afterglow counterparts endowed with sufficient completeness in circum-merger density estimates, one can determine the apparent fraction of high-density mergers. Starting from this number, by estimating the amplification factor related to the high-density-selection effect from population models,} one can constrain the intrinsic fraction of mergers in high-density media. This is the principle of the \new{new} method we \new{propose} in order to study the class of high-density mergers.

As we develop later on, this method should allow \neww{us} to constrain the number of high-density mergers, even after a small number of GW events with afterglow counterpart. The exact link between the rate of high-density events and the distribution of delay times \new{and kick velocities} is not clear, in particular because of the aforementioned uncertainty on the \new{supernova kicks}. Nonetheless, the method we suggest here is a first step toward \new{studying} the delay-time distribution of BNSs from their merger afterglows, \new{at least} for the fastest merging \new{or low-kick binaries}.

This article is organized as follows. In Sect.~\ref{sec:evidence}, we explain the  motivation for developing this new method of constraining high-density mergers by recalling some related observational and theoretical knowledge. In Sect.~\ref{sec:identifying}, we show that, for future BNS merger afterglows, multi-messenger observations will allow \neww{the circum-merger density to be estimated} and, thus, will enable the apparent fraction of high-density mergers to be determined quite accurately. In Sect.~\ref{sec:stats}, we describe how these observations provide significant constraints on the population of high-density mergers even with a limited sample of afterglows, exploiting the sensitivity of the afterglow flux to the circum-merger density. Finally, in Sect.~\ref{sec:discussion} we discuss the limitations of this method, and conclude in Sect.~\ref{sec:conclusion}.

\section{Indirect evidence regarding BNS mergers in dense media}
\label{sec:evidence}

Theoretically, mechanisms exist that lead to fast-merging or low-kick systems. Among these are (i) an efficient common envelope phase, that reduces initial separation \cite[e.g.,][]{ivanova03, dominik+12} and merger time, (ii) a favorable supernova kick, that causes high eccentricity and thus rapid merger or a small migration velocity \citep[e.g.,][]{K96}, \new{(iii) the formation of the BNS by dynamical capture in a migration trap within an active galactic nucleus disk \citep{secunda+19}, or (iv) the interaction of the BNS with another compact object therein \citep{181, fernandez+19}. The frequency with which these actually occur is still unclear.}

Over the years, a body of indirect evidence on high-density mergers has emerged. However, as we show here, current data is inconclusive regarding the \new{importance} of this class of mergers.

First, some population synthesis studies suggest the existence of a `fast' channel for BNS mergers, and, thus, a delay-time distribution featuring a peak around time-scales as short as 20~Myr \citep{pb02,ivanova03,belczynski+06}. These correspond to tight binaries that undergo a third mass transfer episode, and merge while still within star-forming regions in dense environments. These conclusions are corroborated by population study predictions on, for example, $r$-process element abundances in the Milky Way \citep{cote+17} or the redshift distribution of short GRBs \citep{davanzo+14}. The two latter studies suggest a delay-time distribution with a slope $\lesssim -1$, favoring a population of fast mergers, \new{and therefore possibly mergers in dense external media}. However, it has been pointed out that the conclusions of population synthesis studies are somewhat sensitive to the assumptions on the physics of the common envelope phase \citep{dominik+12} or the distribution of natal kicks \citep{safarzadeh+17}.

A second approach is the study of the delay times and kick \new{velocities} of Galactic \new{systems}. This approach is limited by statistics and by the uncertainty in estimating these from observations. However, finding short delay times or weak natal kicks can imply that a significant fraction of double neutron star mergers should occur in regions where star formation may still be significant, and in turn, the densities are large too. Recently, \cite{BP2019} have shown that at least $10-20\%$ of Galactic systems are born with delay times of less than 100~Myr between formation and merger. Furthermore, \cite{BP2016} have shown that the majority of the observed BNSs received relatively weak kicks at birth ($v_{\rm kick} \lesssim 30~{\rm km/s}$, see also \citealt{Tauris2017}).

Another approach is to consider the nature of short GRB host galaxies. On the one hand, these are found to be star-forming two to three times more often than they are found to be elliptical galaxies \citep{Berger2014}. This suggests higher density media for a significant fraction of mergers. This is particularly noteworthy since up to a redshift $z\lesssim 1$, that is, where short GRB hosts can be seen, elliptical and star-forming galaxies share roughly equal fractions of the cosmic stellar mass \citep{Bell2003}. This suggests that short GRBs are preferentially found in lower mass galaxies, and thus experience larger external densities \new{on average} \citep{Zheng2007}. 

Also, the observed host galaxy offset distribution has a median value of 1.5 half-light radii, with $\sim 20\%$ of objects lying outside five half-light radii and $\sim 20\%$ within one half-light radius \citep{fong+13,Berger2014}. This favors higher density environments for the most centered $\sim 20 \%$ of systems. However, host-galaxy completeness of typical samples is small. Moreover, the offset distribution relies on a correct identification of the host galaxy, and may be grossly overestimating the true offset if, for example, the true host is a fainter, unobserved galaxy of lower mass or higher redshift (e.g., \citealt{Behroozi2014}).

Insight into short GRBs occurring in dense environments also comes from GRB afterglow observations. On the one hand, \cite{Nysewander2009} have shown that (i) short and long GRBs present a similar correlation between X-ray flux and gamma-ray fluence, (ii) above a gamma-ray fluence threshold of $10^{-7}\mbox{ erg cm}^{-2}$, optical afterglows are detected in almost all short GRBs and (iii) short and long GRB afterglows have similar radio-to-X-ray flux ratios. These results prompted \cite{Nysewander2009} to suggest that short GRBs have similar or larger external densities to long GRBs, with typical values that may be as large as $1\mbox{ cm}^{-3}$. \neww{For a selected sample of short GRB early afterglows, \citet{brendan} have found that less than 16\% of events took place at densities smaller than $10^{-4}~{\rm cm}^{-3}$, suggesting that few short GRBs occur in very rarefied media.} On the other hand, short GRB afterglow catalogs such as \cite{118} or \cite{Berger2014} do not exhibit a population of high-density afterglows. Similarly, these studies are limited by poor afterglow sampling, parameter degeneracy in photometry fitting and, often, by a lack of the synchrotron self-Compton cooling component in the radiation modeling. In recent years, with the detection of long-lived emission from GRBs with the Large Area Telescope onboard Fermi \citep{A18}, the synchrotron self-Compton cooling channel has been realized to be an important ingredient of the physical picture. \neww{As the Compton parameter affects the position of the cooling frequency, using the cooling break in the X-ray band to estimate the density while disregarding the synchrotron self-Compton effect can particularly bias the result \citep{Beniamini2015}}. These caveats may impede a reliable estimation of the circum-burst density and explain this apparent contradiction.

Finally, an independent approach to short merger binaries comes from $r$-process abundance studies. The arguments in favor of short merger times, \new{and therefore possibly mergers in dense environments,} have recently been summarized in some detail in \citet{HBP2018} and \citet{BP2019}. A prevalence of short merger times is implied by (i) observations of $r$-process enriched stars in ultra-faint dwarf galaxies \citep{BHP2016}, (ii) the large scatter of $r$-process abundances in extremely metal-poor stars in the Milky Way halo \citep{Argast2004,Tsujimoto2014,Wehmeyer2015,Vangioni2016,BDS2018}, (iii) the declining rate of deposition of radioactive $^{244}$Pu and $^{247}$Cm on Earth \citep{Hotokezaka2015,Wallner2015,BH2020} and (iv) the declining rate of [Eu/Fe] as a function of [Fe/H] observed in Milky Way stars for $\mbox{[Fe/H]}\gtrsim -1$ \citep{Matteucci2014,Cote2016,Komiya2016,HBP2018,Simonetti2019}. However, these conclusions rely on knowledge of the rates and $r$-process yields of BNS mergers, core-collapse and thermonuclear supernovae, all of which are still a matter of debate (see \citealt{Cowan19} and \citealt{HBP2018} for reviews respectively on the $r$-process in general and on BNS mergers as its astrophysical site).

\begin{figure}[h]
\includegraphics[width=\linewidth]{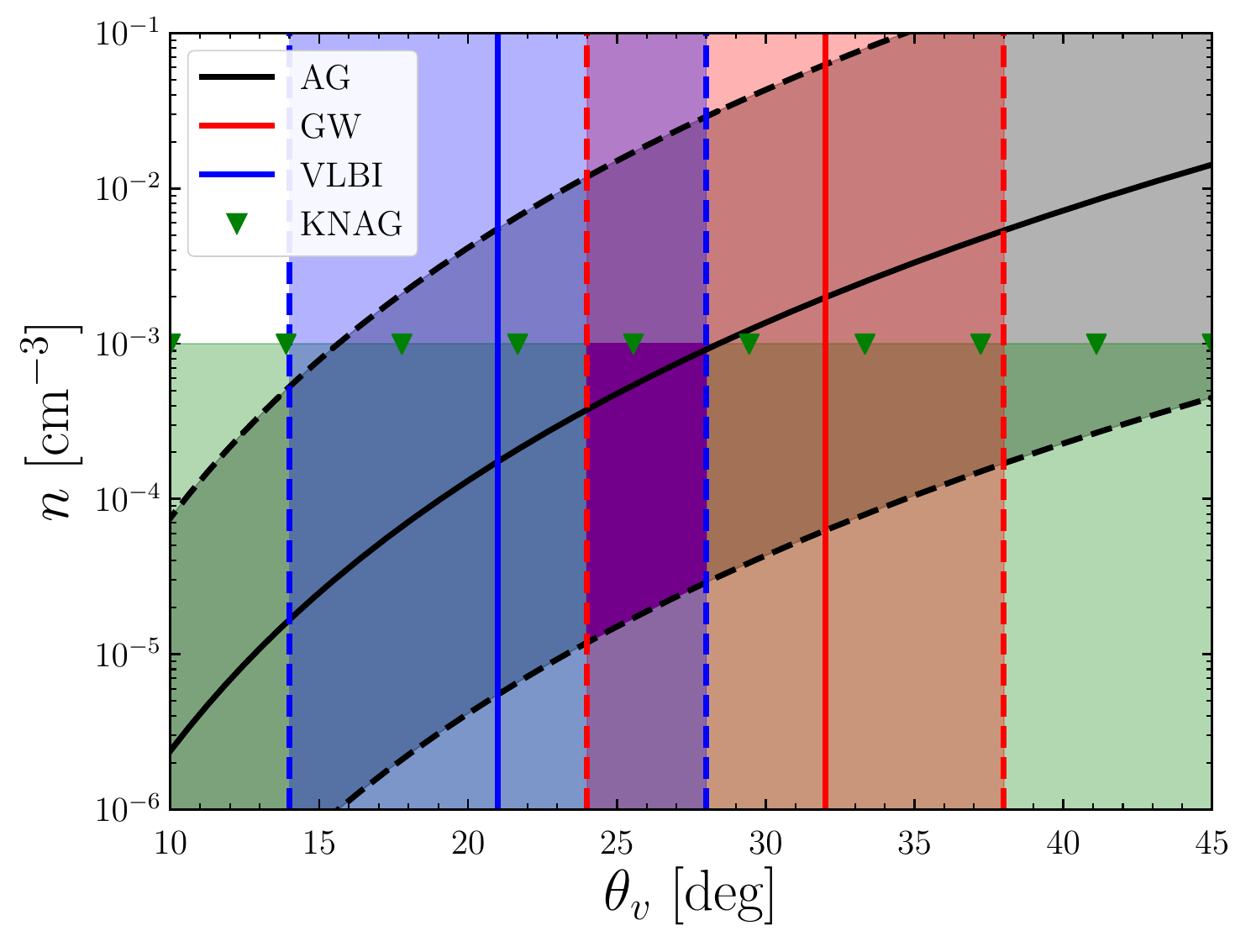}
\caption{Multi-messenger determination of the viewing angle $\theta_v$ and circum-merger medium density $n$ in the case of GRB170817A. We present 1-$\sigma$ confidence regions (solid line: median; dashed line: 68\% confidence limits) obtained from the GW data assuming the source localization (red), the radio afterglow's properties around its peak (black, see Eq.~\ref{eq:tv}) and very long baseline interferometry imaging measurements (blue). Green triangles show the upper limit on $n$ deduced from the (as yet) undetected kilonova afterglow. The preferred region for $\theta_v$ and $n$ is highlighted in purple. The text gives details and references.}
\label{fig:thetavn}
\end{figure}

\section{Determining the apparent fraction of high-density mergers from afterglow observations}
\label{sec:identifying}

We now describe the method we suggest to directly probe the class of high-density mergers. Our method relies on a sample of afterglow counterparts to GW signals from BNS mergers, which would have a sufficient \new{completeness in density} above a certain limiting afterglow flux. Population models such as \cite{145} or \cite{185} apply criteria based on afterglow flux levels, and thus provide predictions on detectable events. Therefore, applying a flux cut to a sample of detected afterglows \neww{ensures} that the sample actually represents all the detectable events above the threshold. This in turn allows one to safely use the predictions from population models to compensate for the density-selection effect and infer the intrinsic fraction of high-density events $f_{\rm HD}$ from the apparent fraction $f_{\rm HD}^{\rm obs}$, that is, the one observed in the sample.

In this section, we describe how to estimate $f^{\rm obs}_{\rm HD}$ for a sample of afterglow counterparts to BNS mergers. This can be done by inferring the densities of individual events from multi-messenger observations, or directly on the level of the entire sample.

\subsection{Measuring the viewing angle and density for a single merger event}
\label{sec:3.1}

Combining the GW and electromagnetic (EM) information channels allows \neww{one} to place individual events quite accurately in the $\theta_v - n$ plane, as has been done in Fig.~\ref{fig:thetavn} for the case of GRB170817A.

First, in Fig.~\ref{fig:thetavn} we present the constraints on $\theta_v$ obtained from the GW data using the information on the event localization from the EM counterpart, as was found by \citet{21}. These are marked in Fig.~\ref{fig:thetavn}, and are representative of three-interferometer constraints that can be obtained in the favorable case where the source is pin-pointed thanks to the detection of the kilonova or early afterglow.

Second, we plot the constraint arising from the properties of the light curve of the radio afterglow around its peak. We start from the equation for the 3~GHz afterglow peak flux $F_p$ and peak time $t_p$ as a function of the jet parameters, in the case where the radio band lies in the $[\nu_m, \nu_c]$ portion of the synchrotron spectrum \citep{114}. \new{Combining these two equations in order to write the ratio $F_p / t_p^3$, we eliminate the jet's kinetic energy from the calculation}. We then insert the equation relating the afterglow peak `shape factor' $\eta = \Delta t / t_2$ to the jet opening and viewing angles \citep{80}. Here, $t_2$ is the onset time of the afterglow's decreasing phase, and $\Delta t$ is the afterglow turnover time, counted between the end of the afterglow's increasing phase and the onset of its decreasing phase. \new{This last operation \neww{eliminates} the jet's opening angle from the calculation, and,} finally, we obtain the following relation between observable quantities (left-hand side) and the jet parameters (right-hand side):

\begin{equation}
\label{eq:tv}
\begin{split}
\left( \frac{F_p}{8.6~{\rm mJy}} \right)\left( \frac{t_p}{4.9~{\rm d}} \right)^{-3} \left( \frac{D}{100~{\rm Mpc}} \right)^{2} \times \left\{ \begin{array}{ll}\left(\alpha \eta \right)^{2} & {\rm no~ex.} \\ 1 & {\rm ex.} \end{array} \right.  & \\
 = \theta_{v, -1}^{-6-2p} n_{-3}^{\frac{p+5}{4}} \epsilon_{e,-1}^{p-1} \epsilon_{B,-3}^{\frac{p+1}{4}}, & 
\end{split}
\end{equation}
where $D$ is the luminosity distance to the event, $\epsilon_e$ and $\epsilon_B$ are the usual shock microphysics parameters\footnote{\neww{These are defined such that a fraction $\epsilon_e$ (resp. $\epsilon_{B}$) of the shocked material's internal energy is carried by the accelerated electron population (resp. the magnetic field).}}, and $\alpha$ is such that the forward shock Lorentz factor is $\Gamma \propto t^{-\alpha}$. For a jet plowing through a uniform medium, $\alpha$ equals 3/8 for a non-expanding jet, and 1/2 for a jet with sound-speed lateral expansion \citep{rhoads99}. The numerical normalization values on the left-hand side of Eq.~\ref{eq:tv} are valid for $p = 2.2$.

We provide these relations in both the expanding and non-expanding jet hypotheses, which are extreme options regarding the jet lateral dynamics. The actual dynamics should lie somewhere in between, and the discrimination between both can be done on the basis of the post-peak afterglow temporal slope (e.g., \citealt{94}). We note that, in the case of an expanding jet, the $\theta_v - n$ relation no longer depends on the turnover time, which may prove difficult to measure in the poorly sampled afterglows of marginally detectable events.

Fortunately, the strongest dependencies here are in the measurable quantities $t_p$, $F_p$ and $D$, rather than on the uncertain $\epsilon_e$ and $\epsilon_B$, allowing \neww{us} to obtain a thin uncertainty region in the $\theta_v - n$ plane. This constraint, \new{which requires only data on the afterglow around its peak}, is shown in Fig.~\ref{fig:thetavn}, where we have taken the values of afterglow observables for GRB170817A from \citet{80}. Here, the width of the uncertainty region is obtained by propagating the 1-$\sigma$ uncertainties on $t_p, F_p, D$ and adding an uncertainty of 0.3 (resp. 2) on $\log \epsilon_{e}$ (resp. $\log \epsilon_{B}$), deduced from the scatter of its value in GRB jet forward shocks \citep{beniamini17,nava14,santana14}. 

Third, we include the viewing angle constraints from the very long baseline interferometry (VLBI) imagery of the radio remnant. By comparing high-resolution imagery of the remnant to synthetic images based on jet models, \citet{79} and \citet{110} were able to constrain the viewing angle to the region shown in blue in Fig.~\ref{fig:thetavn}.

Finally, we add the constraint that comes from the non-detection of the so-called `kilonova afterglow'. This is expected radiation from the forward shock formed by the mildly relativistic material responsible for the kilonova signal on the external medium \citep{82, 83, 188}. Due to the small Lorentz factor and smooth velocity structure of this ejecta, this afterglow component is expected to peak within a decade \new{post-merger} in the case of GRB170817A \citep{188}. The absence of rebrightening of the afterglow, interpreted as the non-detection of the emergence of this component two years after the merger, already constrains the density to $n \lesssim 10^{-3}~{\rm cm}^{-3}$ (\citealt{188}, Fig.~3).

In addition, a detection of the kilonova afterglow would allow for an actual measurement of the density, and not only an upper limit. However, we note that, in both cases, the constraint depends on the assumed value for $\epsilon_e$ in the corresponding shock, which is still uncertain for mildly relativistic shocks. Allowing this parameter to assume values suggested for such shocks  ($\epsilon_e \lesssim 10^{-2}$, \citealt{138}) \new{by particle-in-cell simulations and by observations of young supernova remnants \citep{morlino+12}} loosens the bound on $n$. Therefore, we advise prudence on the use of the kilonova afterglow for measurements of the density. More details on this last point may be found in Sect.~\ref{sec:discussion}.

\begin{figure}
\resizebox{\hsize}{!}{\includegraphics{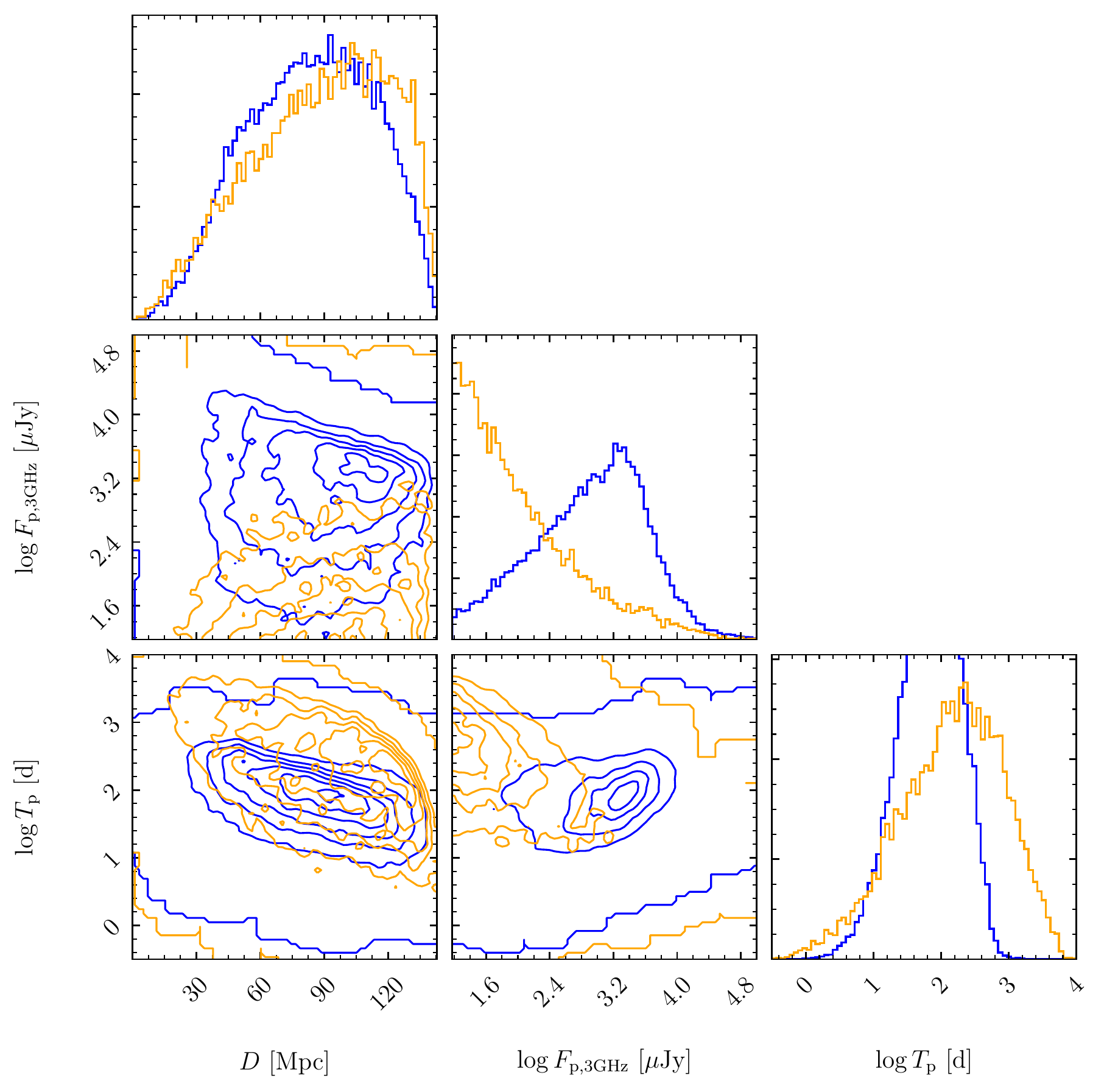}}
\caption{Corner plot of luminosity distance, 3~GHz afterglow peak flux and time of peak of two populations of mergers: one in density of $10^{-3}~{\rm cm}^{-3}$ (yellow), and another in $1~{\rm cm}^{-3}$ (blue). Shown here are synthetic populations for radio-GW jointly detectable events as expected from the population model of \citet{185} for the current O3 run and taking the Very Large Array as the limiting radio instrument, with a 3~GHz sensitivity of 15~$\mu$Jy.}
\label{fig:corner}
\end{figure}

As seen in Fig.~\ref{fig:thetavn}, the combination of the constraints from the GW, the afterglow light curve and VLBI measurement and the kilonova afterglow leads to $\theta_v \in [24,28]^\circ$ and $\log n /{\rm cm}^{-3} \in [-5,-3]$ (all 1-$\sigma$ confidence intervals) for GRB170817A. \new{Disregarding the kilonova afterglow constraint because of the aforementioned uncertainty on $\epsilon_e$ in the corresponding shock, the range of inferred densities becomes $\log n /{\rm cm}^{-3} \in [-5,-2]$.}

Such a combination of constraints is only obtained if all the possible multi-messenger observations are made. Using these after a number of events, an estimate of $f_{{\rm HD}}^{\rm obs}$ can be obtained. It is clear from Fig.~\ref{fig:thetavn} that GW and VLBI data crucially narrow down the constraint on $\theta_v$. Unfortunately, VLBI remnant imagery will likely become impossible in most cases as the GW horizon increases and we expect its contribution to vanish for most events as of the start of the O3 run \citep{185}. In the future, this may be compensated for by some improvement in the GW constraint as more interferometers come online, though it will probably be modest \citep{190, 189}.

\begin{figure*}[ht]
\resizebox{\hsize}{!}{\includegraphics{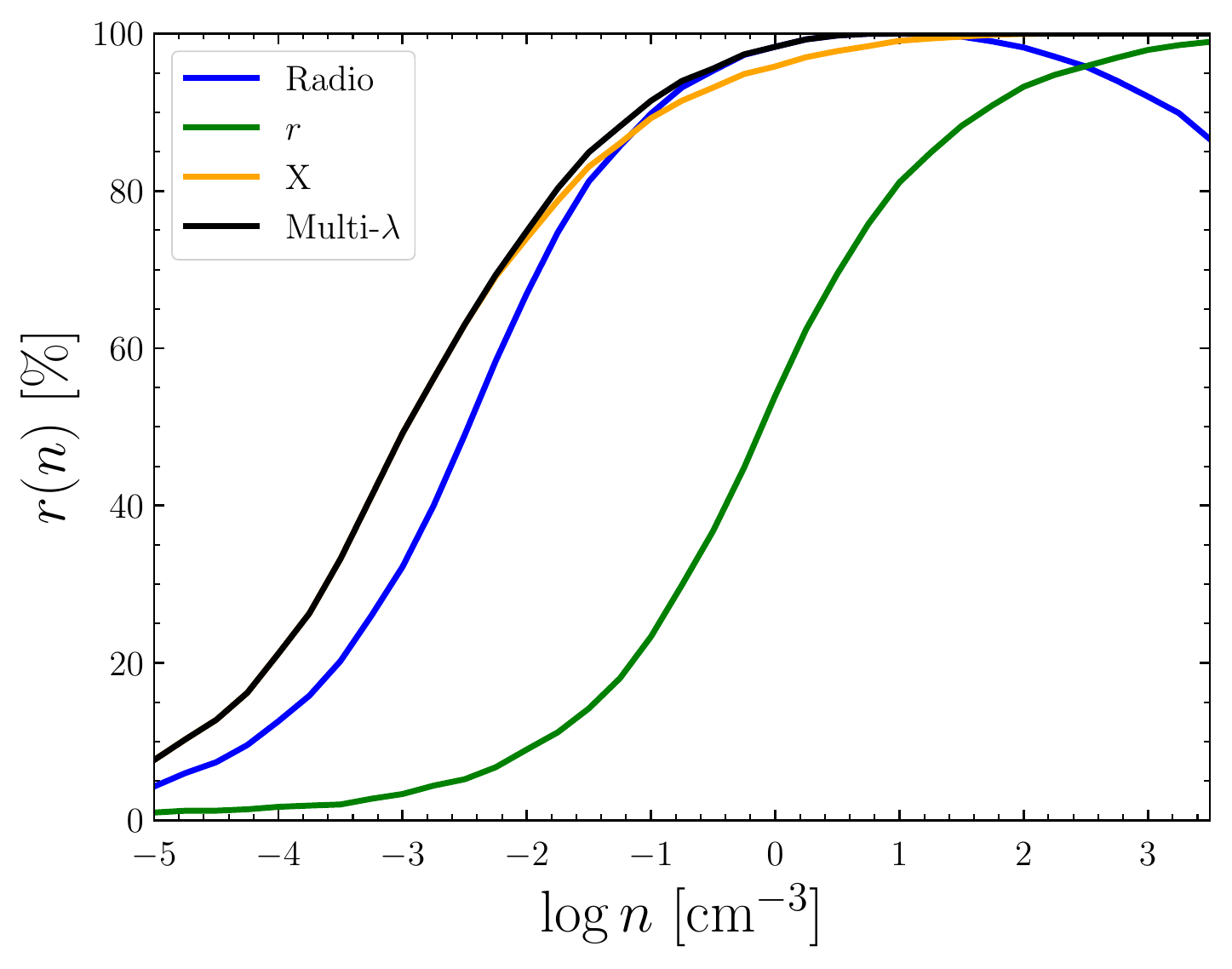}\includegraphics{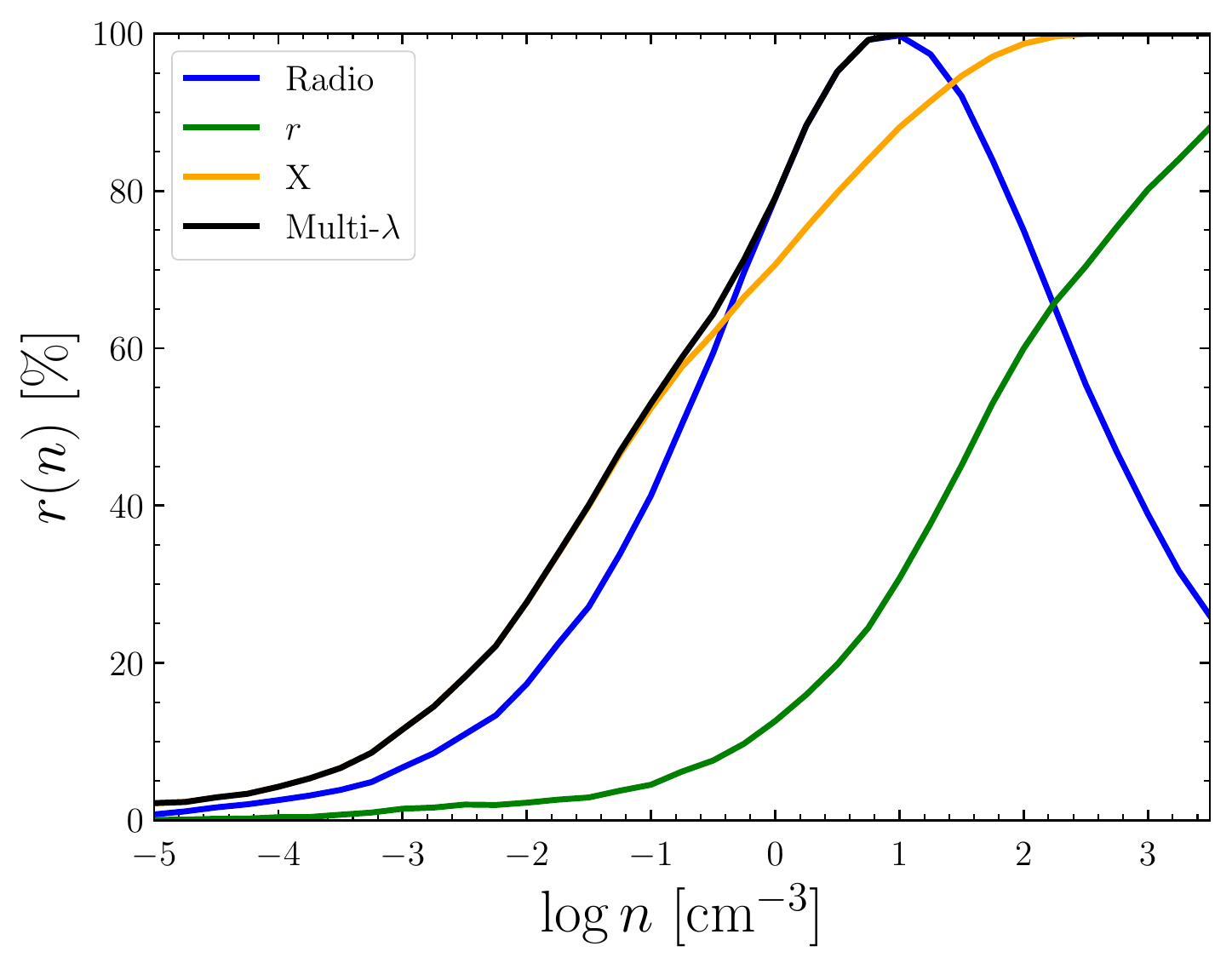}}
\caption{Afterglow recovery fraction in X-ray, optical, and radio bands as function of circum-merger medium density, for a population with energy distribution function deduced from G16 (left) or WP15 (right). We note the effect of synchrotron self-absorption on the recovery fraction in the radio band as of $n \gtrsim 10~{\rm cm}^{-3}$.}
\label{fig:frac}
\end{figure*}

An advantage of this multi-messenger estimation of $n$ is the use of Eq.~\ref{eq:tv}, which requires the properties of the radio afterglow around its peak only and thus is applicable even for faint or poorly-sampled afterglows. Also, it can easily be adapted to other bands, such as the optical, provided they lie between $\nu_m$ and $\nu_c$ \new{and the afterglow is not outshined by the kilonova}. However, Eq.~\ref{eq:tv} is valid only for small densities, when the effects of synchrotron self-absorption in the forward shock are negligible. As illustrated later in Fig.~\ref{fig:frac}, this is no longer the case as soon as $n \gtrsim 10-100~{\rm cm}^{-3}$, depending on the distribution of jet kinetic energies of the population. Nonetheless, from Fig.~\ref{fig:frac}, one expects that at these densities, the X-ray afterglow will be readily accessible and $n$ can be estimated from fully-fledged afterglow fitting, containing more physics than Eq.~\ref{eq:tv}.

\subsection{Using $n$ -- $\theta_v$ correlations in the sample of merger afterglows}

If such follow-up observations are not done and the only available data are GW and afterglow photometry, $f_{{\rm HD}}^{\rm obs}$ can still be retrieved at the level of the observed sample thanks to important density-dependent correlations in the afterglow peak properties.

In Fig.~\ref{fig:corner}, we plot the distributions of the distance, 3~GHz afterglow peak flux and peak time for two populations of mergers, in high- or low-density media. These are the distributions for the mergers predicted to be detectable \neww{jointly in GW and in the radio band} by the VLA (with a limiting sensitivity of 15~$\mu$Jy) for the O3 run \neww{and supposing they are} placed in media with unique high ($n = 1~{\rm cm}^{-3}$) or low ($n = 10^{-3}~{\rm cm}^{-3}$) densities. \neww{These distributions, as all the afterglow populations mentioned throughout this article, were generated as in \citet{185}. That is, progenitor binaries were assumed uniform in space within the GW horizon and isotropic in jet direction, which we suppose is the direction of the system's angular momentum. For each binary, the jet's energy was sampled from an energy distribution function (in Fig.~\ref{fig:corner}, this was deduced from \citealt{104}, see details in Sect.~\ref{sec:stats}) and the afterglow radiation was computed using the full synchrotron spectrum, including self-absorption, in the thin shell regime, supposing a top-hat jet with relativistic deceleration dynamics. In the sample, events were deemed GW- and radio-detectable by applying thresholds on their GW signal-to-noise ratio and afterglow peak flux, respectively}. Synchrotron self-Compton effects were ignored in this analysis, as frequencies are always well below $\nu_c$.

In particular for $t_p$ and $F_p$, the distributions are qualitatively different. The low-density mergers accumulate around the limiting flux, showing that the bulk of the population is undetectable, whereas the high-density mergers present a peak at the mJy level. The combination of these population-level correlations with an adequate statistical treatment of afterglow observations should allow \neww{one} to estimate $f_{{\rm HD}}^{\rm obs}$ for the sample.

\section{Constraining high-density mergers with $f_{{\rm HD}}^{\rm obs}$}
\label{sec:stats}

\begin{figure*}[ht]
\resizebox{\hsize}{!}{\includegraphics{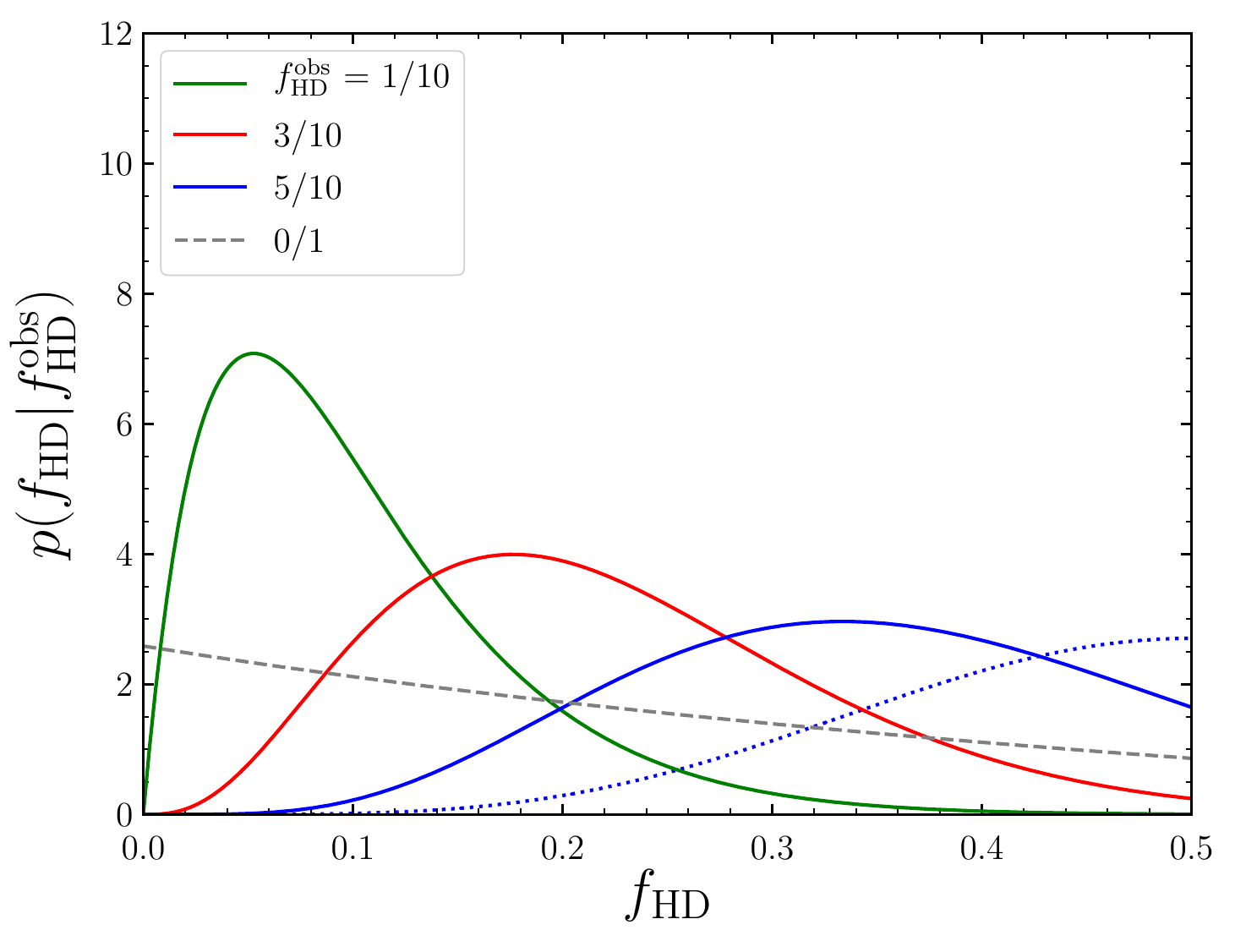}\includegraphics{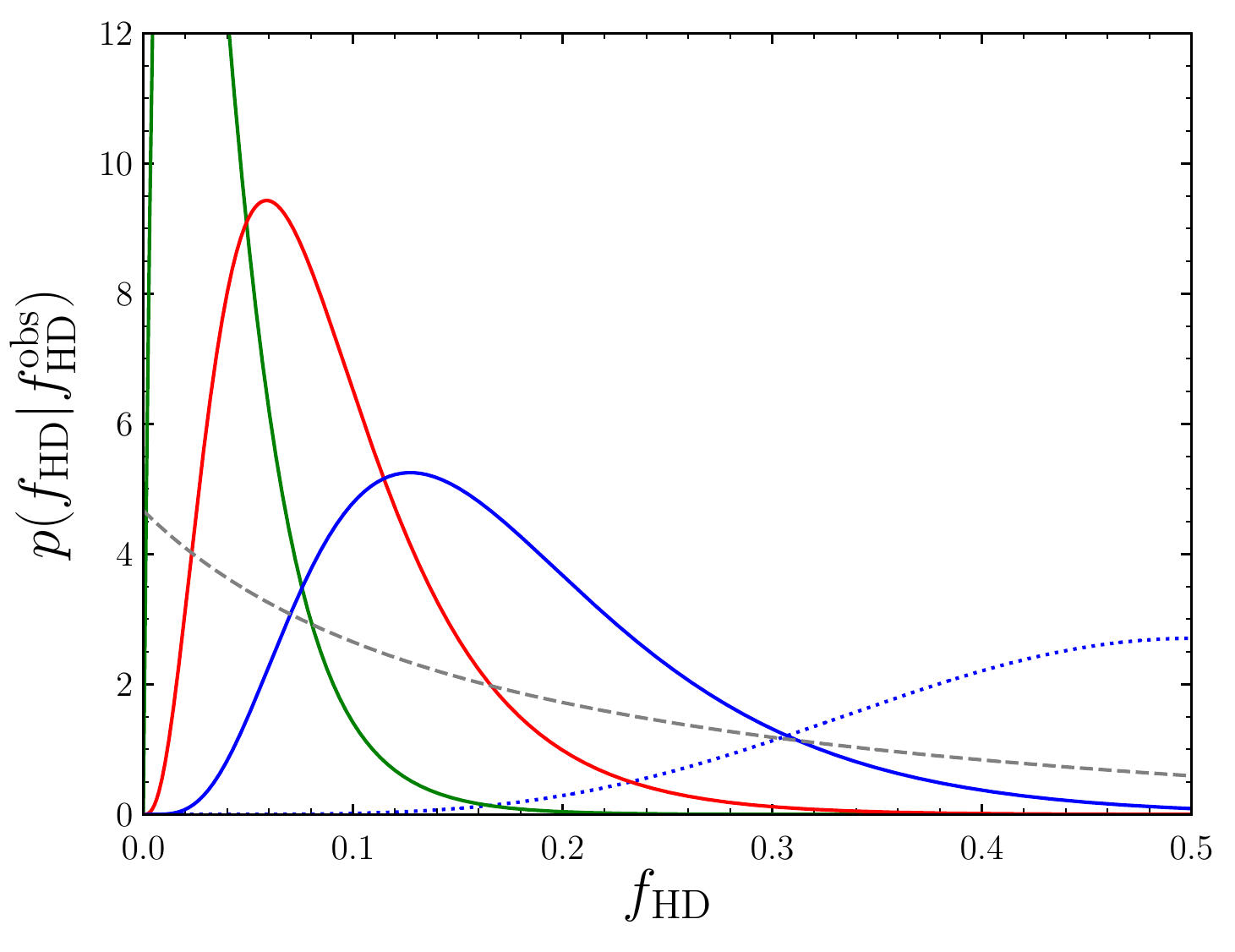}}
\caption{Posterior probability density function of $f_{{\rm HD}}$ obtained after having observed fraction $f_{{\rm HD}}^{\rm obs}$ of high-density mergers among ten events, for varying $f_{{\rm HD}}^{\rm obs}$. The dashed line shows the current constraint, obtained after the single low-density event GRB170817A. The dotted blue line shows the constraint obtained with $f_{{\rm HD}}^{\rm obs} = 5/10$, but ignoring the selection effect, i.e., with $r(n_1) = r(n_2) = 1$. Left: Assuming the population's jet kinetic energy distribution follows the short GRB luminosity function of G16. Right: Same, for that of WP15.}
\label{fig:fig2}
\end{figure*}

We now illustrate our method of constraining high-density mergers starting from their apparent fraction $f_{{\rm HD}}^{\rm obs}$ obtained from multi-messenger follow-up campaigns, as shown in Sect.~\ref{sec:identifying}.

For the sake of illustration, suppose mergers occur in two different types of media: high-density ($n_2$) and low-density ($n_1 \leq n_2$). We are interested in inferring from multi-messenger BNS merger observations the intrinsic fractions $f_{{\rm HD}}$ and $f_{\rm LD} = 1 - f_{{\rm HD}}$ of mergers occurring respectively in media of densities $n_2$ and $n_1$.

For a certain EM band $B$, let $r_B(n)$ denote the `afterglow recovery fraction' at density $n$, meaning the fraction of mergers occurring at density $n$ to produce a detectable afterglow in the $B$ band. This is provided in Fig.~\ref{fig:frac} for the X-ray (1 keV), optical ($r$), and radio (3~GHz) bands, assuming detection limits respectively of $10^{-15}~{\rm erg/s/cm}^2$ (50~ks exposure of Chandra in 0.5-8~keV band), magnitude 24 (space telescope routine observation) and $15~\mu {\rm Jy}$ (18~ks exposure of VLA in 2-4~GHz band). The plotted $r_B(n)$ were determined from populations synthesized for the O3 run as in Fig.~\ref{fig:corner}, but placed in media with densities that are constant within a population but varying from one population to another. Furthermore, they assume two different distributions for the jet kinetic energies: one deduced from the short GRB luminosity function of \citet{105} (hereafter, WP15), the other from that of \citet{104} (hereafter, G16). In both cases, \neww{we deduced} the jet kinetic energy distribution from the short GRB luminosity function assuming typical short GRB durations and $\gamma$ efficiencies of 0.2~s and 20\%. Also, we give the multiwavelength afterglow recovery fraction $r_{{\rm M}\lambda}(n)$, which accounts for events detectable in at least one of the three bands.

\neww{The luminosity functions found in G16 and WP15 were deduced from distinct GRB data sets and using methods with distinct hypotheses. Among GRB luminosity function studies, they represent two extremes in terms of population steepness, with WP15's luminosity function being much more bottom-heavy than that of G16. For our purposes, G16 can be understood as optimistic with regards to afterglow detectability, and WP15 pessimistic.} 

As mentioned in Sect.~\ref{sec:3.1}, synchrotron self-absorption tends to decrease $r_{3{\rm GHz}}(n)$ as of $n \gtrsim 10-100~{\rm cm}^{-3}$, which appears clearly in Fig.~\ref{fig:frac}. This leads us to consider other bands (and most prominently the X-ray) for the estimation of $n$ in individual events. Therefore, we shall consider $r_{{\rm M}\lambda}$ as the relevant recovery fraction in what follows.

As explained in Sect.~\ref{sec:intro}, because of the strong dependence of the afterglow peak flux to the circum-merger density ($F_p \propto n^{\frac{p+1}{4}}$), we have $r(n_1) \ll r(n_2)$. Therefore, mergers in high-density media should be over-represented in the observed population with respect to their intrinsic fraction $f_{{\rm HD}}$. This establishes a method to effectively constrain the latter following the observation of only a few of these high-density events.

The probability of observing a high-density merger is 
\begin{equation}
p_{{\rm HD}} = \frac{r(n_2) f_{{\rm HD}}}{r(n_1)f_{\rm LD} + r(n_2)f_{{\rm HD}}}.
\end{equation}

Furthermore, after observing $N$ afterglow counterparts to GW, the likelihood that a fraction $f_{{\rm HD}}^{\rm obs}$ will be found to occur in a high-density medium is that of a binomial process with success probability $p_{{\rm HD}}$ and $N$ tries\footnote{\neww{Here we denote the binomial coefficient ${a \choose{b}} = \frac{b!}{a!(b-a)!}$.}}:
\begin{equation}
p(f_{{\rm HD}}^{\rm obs} | f_{{\rm HD}}, N) = {N \choose{f_{{\rm HD}}^{\rm obs} N}}\, p_{{\rm HD}}^{f_{{\rm HD}}^{\rm obs} N} \left(1 - p_{{\rm HD}}\right)^{(1-f_{{\rm HD}}^{\rm obs}) N}.
\end{equation}

Finally, since according to Bayes' theorem with no prior information on $f_{{\rm HD}}$ we have $p(f_{{\rm HD}} | f_{{\rm HD}}^{\rm obs}, N) \propto p(f_{{\rm HD}}^{\rm obs} | f_{{\rm HD}}, N)$, a constraint on $f_{{\rm HD}}$ follows. Given the high sensitivity of the fraction $r(n)$ to the density, we expect these constraints to be tight even with a small number of events.

This is clear in Fig.~\ref{fig:fig2}, where we have chosen $n_1 = 10^{-3}~{\rm cm}^{-3}$, $n_2 = 1~{\rm cm}^{-3}$, and we show the constraints that could be obtained from ten events (as expected after three years of an O3-type run, \citealt{185}) among which one, three or five are in a high-density medium. We observe that the constraints do not center around $f_{{\rm HD}}^{\rm obs}$ and are tighter than if the bias towards high-density events were ignored, as can be seen by comparing the solid blue curves with the dotted blue curves. This illustrates the `magnifying effect' of the selection by the afterglow.

The slope of the jet energy function is steeper for WP15 than for G16. This implies that, overall, G16 predicts more high-energy events than WP15. This explains why $r(n)$ is systematically larger for G16 than for WP15, at least in the regime where $F_p \propto E n^{\frac{p+1}{4}}$, that is, before the onset of the self-absorption suppression. This also implies that the rate at which afterglows are recovered by increasing the density is greater for WP15 than for G16. In terms of recovery fraction, this is expressed by saying that the contrast $r(n_2)/r(n_1)$ is larger for WP15 than for G16, which naturally leads to tighter constraints, as is clear from Fig.~\ref{fig:fig2}.

In the case where no high-density events are observed, upper limits on the intrinsic fraction $f_{{\rm HD}}$ can be deduced. This is done in Table~\ref{tab:limits}, where we report the 95\%-confidence level upper limits deduced from the observation of $N$ events, all in low-density media. It appears that the observation of only five low-density events (e.g., observing exclusively low-density events during 18 months of an O3-type run, \citealt{185}) suffices to constrain $f_{{\rm HD}}$, at the 95\%-confidence level, to being smaller than 18.5\% (resp. 9.4\%), assuming the short GRB luminosity function of G16 (resp. WP15).

\begin{table}
\caption{The 95\%-confidence level upper limits on $f_{\rm HD}$ obtained after observing no high-density events among $N$ afterglows, in two short GRB energy function distribution hypotheses.}
\label{tab:limits}
\begin{tabular}{llllll}
\hline
\hline
$N$ & 1 & 5     & 10  & 20 & 50 \\
\hline
G16 & 70 \%& 18.5 \%& 8.5 \%& 4 \% & 1.5\% \\
WP15 & 64 \%& 9.4\% & 3.9\% & 1.7\% & 0.7\% \\
\hline
\end{tabular}
\end{table}

\section{Discussion}
\label{sec:discussion}

We have presented a method of effectively constraining the class of BNS mergers that occur in high-density media. It is based on the observation of their afterglow counterparts. We will now discuss the limitations, conditions for application and possible extensions of this method.

\subsection{Method limitations and applicability}

A first limitation of the method presented here is the requirement that the sample be density-complete above a certain afterglow flux. In other words, it requires the \neww{certitude} that all detectable afterglows with fluxes above a limit were effectively detected. Only in this case can the model-determined recovery fraction $r(n)$ be used to infer $f_{{\rm HD}}$ from $f^{\rm obs}_{{\rm HD}}$. As the observational biases resulting in practical limitations to these detections are discussed in \citet{185}, we do not repeat them here. We only mention that the difficulty in following-up GW events linked to the size of the localization sky-maps should be met by large-field facilities such as the Zwicky Transient Facility \citep{ZTF}, and by future high-cadence survey instruments such as the Large Synoptic Survey Telescope \citep{LSST}. In practice, density-completeness will be difficult to obtain, and an uncertainty on $f^{\rm obs}_{{\rm HD}}$ must be taken into account in applying this method.

Furthermore, there is a selection bias towards high-density mergers for reasons unrelated to the afterglow flux. For instance, afterglows of mergers occurring in denser media should peak at earlier times, favoring their detection during follow-up, regardless of their flux level. Consequently, the flux-related selection bias we quantified here in $r(n)$ actually underestimates the bias toward high-density events.

Similarly, there is a selection bias towards bright afterglows regardless of the events' circum-merger density. For instance, events closer or brighter in gamma-rays should be better localized by the GW or GRB data, easing their follow-up, regardless of the circum-merger density. These density-unrelated biases towards afterglow detection actually correlate positively with afterglow flux and thus, statistically, with density. Therefore, once again, the bias towards high-density events we consider here is underestimated.

This method is not applicable to the population of cosmological short GRBs for which densities have been estimated, for two main reasons. The first is that the densities claimed for this population are deduced from uncertain fits, as argued in Sect.~\ref{sec:evidence}, and that only a small fraction of GRBs have a claimed density. Thus, the resulting $f^{\rm obs}_{{\rm HD}}$ would be quite uncertain. The second is that, for these regular short GRBs, the afterglow detectability depends more on factors that are not density-related, such as (i) the availability of sufficiently rapid follow-up observations and other human factors, or (ii) the quality of the localization of the GRB, which is linked to its prompt properties and not to its afterglow. Also, for regular GRBs, the expected recovery fraction $r(n)$ should be determined through a population model selecting events on joint GRB-afterglow detection, instead of on joint GW-afterglow.

\new{In Sect.~\ref{sec:identifying}, we mentioned the kilonova afterglow as an alternative means of measuring the merger environment density regardless of the viewing angle, as allowed by the quasi-isotropy of this signal. Nonetheless, we caution against the feasibility and robustness of such a measurement. First of all, as shown by particle-in-cell simulations and the observation of young supernova remnants \citep{138, morlino+12}, mildly relativistic shocks are expected to be poor electron accelerators, with $\epsilon_e$ up to orders of magnitude lower than in relativistic shocks. Therefore, seeing as the afterglow flux scales with $\epsilon_e$, the kilonova afterglow should be significantly fainter than the jet's afterglow and unlikely to be detectable in most cases. Furthermore, in the typical case of a low-density medium, this signal is expected to peak up to a decade post-merger, posing some challenge to its detection in follow-up campaigns. Finally, the kilonova afterglow light curve depends on the minimal velocity of the merger ejecta and on its radial velocity structure, both of which are still uncertain for lack of modelling and observation history. Therefore, although the kilonova afterglow signal's quasi-isotropy dismisses the degeneracy between the density and the viewing angle, its use introduces some uncertainty to the measurement, which is thus rendered not robust.}

\subsection{From mergers in dense environments to fast-merging and low-kick binaries}

In Sect.~\ref{sec:intro}, we presented this method of determining $f_{{\rm HD}}$ as a first step towards constraining the population of fast-merging binaries required to explain various astrophysical data, as summed-up in Sect.~\ref{sec:evidence}.

First of all, an astrophysically interesting constraint on the densities of circum-merger media should be given as a continuous distribution of densities within the population, and not only as a fraction of high- and low-density mergers as we have shown here for simplicity. A continuous (parametric) distribution of densities does not pose any mathematical problems and can be included in this method.

Second, constraining the distribution of merger delay times from that of the merger \new{environment} densities is non-trivial, because the medium hosting the merger \new{effectively} depends on the locus of the second supernova \new{in the galaxy}, on the kick it imparts to the binary system, on the galactic potential, and on the galactic density profile. All of these are uncertain or variable from system to system. As stated in Sect.~\ref{sec:evidence}, efforts to tackle these effects on the level of population-synthesis models have been done, and are ongoing (O'Connor et al., in prep). Nonetheless, untangling the effects of all these factors remains difficult.

Third, our method relies on observing the afterglow counterparts to GW inspiral signals, and thus can only inform us on the high-density mergers within the horizon of the GW instruments. However, the fast-merging binary population suggested by the $r$-process element observations mentioned in Sect.~\ref{sec:evidence} must have formed and enriched their hosts shortly after the peak of cosmic star formation, that is, at $z \sim 2$. Thus, this method will remain ineffective with regards to this particular population, as long as we rely on second-generation GW instruments. However, with the prospect of detecting inspiral signals from systems at $z \gtrsim 1$ with third-generation interferometers such as the Einstein Telescope \citep{ET} or Cosmic Explorer \citep{CE}, the constraining power of this method becomes larger and extends to the redshifts where fast-merging binaries are a matter of debate. In this context, a complete description should require a redshift-varying fraction $f_{{\rm HD}}(z)$, the addition of which is a straightforward extension of our method. \new{At these redshifts, however, detection of the kilonova may reveal challenging and the localization of the source needed for multiwavelength follow-up should be \neww{ensured} directly by detection of the afterglow by wide-field X-ray instruments such as Theseus \citep{theseus} or radio survey facilities such as the Square Kilometer Array \citep{SKA}.}

\section{Conclusion}
\label{sec:conclusion}
We have described a method of directly probing the binary neutron stars that merge in \new{dense} environments, based on the observation of binary neutron star merger afterglows and exploiting the high sensitivity of these to the circum-merger medium density. Its constraining power is large and, since high-density mergers are naturally associated with fast-merging \new{or low-kick} binaries, this method is a first step toward a new independent approach to the delay-time \new{and kick velocity} distributions.

\section*{Acknowledgments}
The authors thank the anonymous referee for constructive remarks, and thank L. Resmi and S. Vergani for useful discussions. R. Duque, F. Daigne and R. Mochkovitch acknowledge financial support from the Centre National d’Études Spatiales (CNES). P. Beniamini's research was funded in part by the Gordon and Betty Moore Foundation through Grant GBMF5076.

\bibliographystyle{aa} % style aa.bst
\bibliography{main}

\end{document}